\documentclass[11pt,letterpaper]{amsart}

\usepackage[latin1]{inputenc}
\usepackage{amsmath}
\usepackage[small,sc]{caption}
\usepackage[dvips]{graphicx}

\newcommand{\cal}[1]{{\mathcal{#1}}}

\newcommand{\btau}{{\bar{\tau}}}
\newcommand{\sigmag}{{{}^3 \! g}}

\begin{document}

\title[Märzke-Wheeler coordinates]{Märzke-Wheeler coordinates for accelerated observers in special relativity}

\author[M.~Pauri and M.~Vallisneri]{M.~Pauri$^1$ \and M.~Vallisneri$^2$}

\date{September 5, 2000}

\begin{abstract}
In special relativity, the definition of coordinate systems adapted to
generic accelerated observers is a long-standing problem, which has
found unequivocal solutions only for the simplest motions.  We show
that the \emph{Märzke-Wheeler construction}, an extension of the
Einstein synchronization convention, produces accelerated systems of
coordinates with desirable properties: (a) they reduce to Lorentz
coordinates in a neighborhood of the observers' world-lines; (b) they
index \emph{continuously} and \emph{completely} the \emph{causal
envelope} of the world-line (that is, the intersection of its causal
past and its causal future: for well-behaved world-lines, the entire
space-time). In particular, Märzke-Wheeler coordinates provide a
smooth and consistent foliation of the causal envelope of any
accelerated observer into space-like surfaces.

We compare the Märzke-Wheeler procedure with other definitions of
accelerated coordinates; we examine it in the special case of
stationary motions, and we provide explicit coordinate transformations
for uniformly accelerated and uniformly rotating observers. Finally,
we employ the notion of \emph{Märzke-Wheeler simultaneity} to clarify
the relativistic paradox of the twins, by pinpointing the local origin
of differential aging.
\end{abstract}

\maketitle

\footnotetext[1]{Dipartimento di Fisica, Universit\`a di Parma, 43100
Parma, Italy; INFN, Sezione di Milano, Gruppo Collegato di Parma,
Italy. E-mail address: pauri@parma.infn.it}

\footnotetext[2]{Theoretical Astrophysics 130-33, Caltech, Pasadena CA
91125; Dipartimento di Fisica, Universit\`a di Parma, 43100 Parma,
Italy; INFN, Sezione di Milano, Gruppo Collegato di Parma,
Italy. E-mail address: vallis@tapir.caltech.edu}

\setcounter{footnote}{2}

\section{Introduction}
\label{sec:introduction}

\noindent In the usual textbook special relativity, the distinction
between ``inertial observer'' and ``Lorentz coordinate frame'' is
blurred. Because of the symmetries of Minkowski space-time, inertial
observers can label all the events of space-time in a simple and
consistent manner that is based on physical conventions and idealized
procedures. (For example, inertial observers can be thought to set up
Lorentz coordinate frames via a framework of ideal clocks and rigid
rods that extend throughout the space-time region of interest,
outfitting it with suitable measuring devices; the clocks are
synchronized with light signals; and so on. See, for instance,
\cite{Misner73}.)

For inertial observers, Lorentz coordinates are a device to extend
their concept of physical reality from their world-line to the entire
space-time, building a description of the world which incorporates
notions of \emph{distance}, and \emph{simultaneity}. What is more,
this description of physics is translated easily between inertial
observers in relative motion with respect to each other, by the
transformations of the Poincaré group.

It follows that in special relativity many physical notions have a
joint local and global valence: they are defined with reference to the
\emph{entire} Minkowski space-time, but they also carry a well-defined
meaning for \emph{local} inertial observers. An instance is the notion
of ``particle'' in quantum field theory (see, {e.\ g.},
\cite{Itzykson85,Pauri99}), which corresponds to a \emph{global},
quantized classical mode of the field extending across Minkowski
space-time, but also to the outcome of \emph{local} detections along
an observer's trajectory.

Now, suppose we are interested in the observations made by
\emph{non-inertial} observers: of course we could study their physics
in some given ``laboratory'' inertial frame of reference. Yet if we
could rewrite all equations in a set of coordinates that is somehow
\emph{adapted and natural} to the observers' accelerated motion, we
would obtain an interesting representation of the ``intrinsic''
physics that the accelerated observers experience and theorize
about. A well-known example is the Unruh effect \cite{Unruh76}, where
``laboratory'' physics predicts that a uniformly accelerated observer
moving through the Minkowski quantum vacuum will behave as if in
contact with a thermal bath, while ``intrinsic'' physics describes the
Minkowski vacuum as consisting of a thermal distribution of quantum
``particles'', as defined by the accelerated observer.

\section{Definition of coordinates for accelerated observers}
\label{sec:definition}

\noindent We set out to define an adapted coordinate system for an
accelerated observer (we shall call him ``Axel'') who is moving
through Minkowski space-time.  Since the accelerated system should
describe Axel's ``intrinsic'' physics, its time coordinate should
coincide with Axel's proper time. Moreover, around any event of Axel's
world-line, there is a small neighborhood where the accelerated
coordinates should approximate the local, instantaneous Lorentz rest
frame. To satisfy these requirements, we can propagate a
\emph{Fermi-Walker transported tetrad}\footnote{ Fermi-Walker
transported vectors ``change from instant to instant by precisely that
amount implied by the change of the four-velocity'' \cite{Misner73};
the transported four-vectors for Axel then obey
\begin{equation}
\frac{dv^\mu}{d\tau} = (u^\mu a^\nu - u^\nu a^\mu)v_\nu,
\end{equation}
where $\tau$ is Axel's proper time, $u^\mu=dx^\mu/d\tau$ his
four-velocity, and $a^\mu = du^\mu/d\tau$ his acceleration.} along
Axel's world-line, and use the tetrad vectors (one of which will point
along Axel's four-velocity) to reach out to Axel's surroundings.

\begin{figure}
\includegraphics{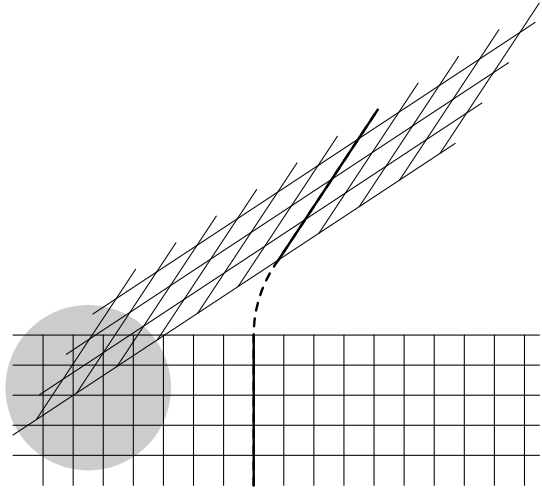}
\caption{World-line of an observer who undergoes a brief period of
acceleration (shown dashed). Extension of the \emph{Fermi-Walker
transported} coordinate system runs into trouble when different
constant-time surfaces overlap on the left. (Adapted from
\cite{Misner73}.)\label{fig:overlap}}
\end{figure}

How do we extend these prescriptions to cover the entire Minkowski
space-time? We can define \emph{extended-tetrad coordinates} by
stretching out rigidly the Fermi-Walker transported axes beyond Axel's
immediate vicinity, but we run into trouble soon: for instance, if
Axel (a) starts at rest, (b) moves for a while with constant
acceleration $|a| = g$, then (c) continues with constant velocity (see
Fig.~\ref{fig:overlap}), we find that the constant-time surfaces of
phase (a) overlap with those of phase (c), at a distance of order
$g^{-1}$ from Axel's world-line. The constant-time planes intersect
because they are orthogonal to Axel's four-velocity $u^\mu$, which
tilts during accelerated motion.

\section{The Märzke-Wheeler procedure}
\label{sec:procedure}

\begin{figure}
\includegraphics{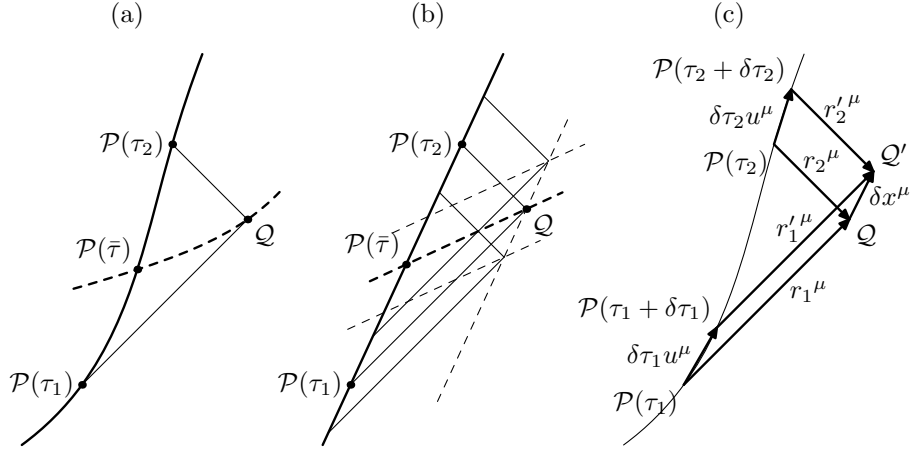}
\caption{Definition of Märzke-Wheeler coordinates. (a): General
case. (b): Inertial case. Märzke-Wheeler coordinates reproduce a
Lorentz frame. (c): Proof that the constant-$\btau$ surfaces are
space-like (see p.\ \pageref{pr:spacelike}).\label{fig:marzke}}
\end{figure}

\noindent We need a way to \emph{foliate} Minkowski space-time into
\emph{non-overlapping surfaces of simultaneity} that are adapted to
Axel's motion and that reduce to local Lorentz frames around his
world-line. Märzke and Wheeler \cite{Marzke64} discussed an extension
of \emph{Einstein's synchronization convention}\footnote{By Einstein's
\emph{convention}, two \emph{inertial} observers get synchronized by
exchanging light signals, while assuming that the one-way speed of light
between the inertial world-lines is equal to the average round-trip
speed. The resulting notion of simultaneity yields the standard slicing
of Minkowski space-time into hyperplanes of constant Lorentz coordinate
time. Since the work of Reichenbach \cite{Reichenbach57} and Gr\"unbaum
\cite{Grunbaum73}, the issue of the conventionality of simultaneity has
generated much contention, mainly on philosophical grounds. Malament
\cite{Malament77} showed that Einstein simultaneity is uniquely
definable from the relation of \emph{causal connectibility}, so it
should be considered \emph{non-conventional} in Gr\"unbaum's sense. On
this point, see also \cite{Sarkar99,Rynasiewicz00}.}
to synchronize observers in curved space-time. The notion of
Märzke-Wheeler simultaneity, \emph{restricted to accelerated observers
in flat space-time}, has just the properties we need\footnote{Our
construction bears resemblance to some applications of Milne's
\emph{k-calculus} \cite{Page36} and to other arguments in the
literature \cite{Ives50,Whitrow61,Kilmister93}.}. We use it to build
\emph{Märzke-Wheeler coordinates}, specified as follows. Imagine that:
(a) at each event along his world-line $\cal{P}(\tau)$, accelerated
observer Axel emits a flash of light imprinted with his proper time;
(b) in the spatial region that Axel wants to monitor, there are
labeling devices capable of receiving Axel's flashes and of sending
them back with their signature; (c) Axel is always on the lookout for
returning signals (see Fig.\ \ref{fig:marzke}a).  Now, suppose that at
event $\cal{Q}$ a labeling device receives and rebroadcasts a light
flash originally emitted by Axel at proper time $\tau_1$, and that
Axel receives the returning signal at proper time $\tau_2$. Then Axel
will \emph{conventionally} label $\cal{Q}$ with a time coordinate
$\bar{\tau} = (\tau_1 + \tau_2)/2$ and a radial coordinate $\sigma =
(\tau_2 - \tau_1)/2$. These two coordinates can then be completed by
two angular coordinates which specify the direction of $\cal{Q}$ with
respect to $\cal{P}(\bar{\tau})$.

If $\cal{P}(\tau)$ is an inertial world-line, the constant-$\btau$
surfaces are just constant Lorentz-time surfaces, and $\sigma$ is
simply the radius of $\cal{Q}$ in spherical Lorentz coordinates (see
Fig.\ \ref{fig:marzke}b): for inertial observers, Märzke-Wheeler
coordinates reduce to Lorentz coordinates (see App.~\ref{app:twins}
for a proof in a special case). Even better, this procedure yields
well-defined coordinates $\bar{\tau}$ and $\sigma$ \emph{for any event
$\cal{Q}$ that lies in the intersection of the causal past and causal
future\footnote{See Wald \cite{Wald84}, ch.~8, for these and other
definitions concerning the causal structure of space-time.} of the
world-line $\cal{P}(\tau)$} (we shall refer to this set as the
\emph{causal envelope} of $\cal{P}(\tau)$; it contains all the events
from which bi-directional communication with Axel is possible). Proof:
(a) the past and future light-cones of $\cal{Q}$ necessarily intersect
with $\cal{P}(\tau)$ somewhere, by definition of causal future and
past; (b) the intersection of a null surface with a time-like curve is
unique, so once $\cal{Q}$ is given, $\tau_1$ and $\tau_2$ are
well-defined. It follows also that constant-$\btau$ surfaces cannot
intersect.

\label{pr:spacelike} We shall use the notation $\Sigma_\btau$ to refer
to the surface of simultaneity labeled by the Märzke-Wheeler time
$\btau$. To prove that each $\Sigma_\btau$ is space-like, refer to
Fig.~\ref{fig:marzke}c, and consider a point $\cal{Q}'$ that is
displaced infinitesimally from $\cal{Q}$; the future light cone with
origin in $\cal{Q}'$ intersects Axel's world-line at the event
$\cal{P}(\tau_2 + \delta \tau_2)$. Define
\begin{equation}
\begin{split}
{r_2}^\mu & \equiv \bigl( \cal{Q} - \cal{P}(\tau_2) \bigr)^\mu, \\
	{r'_2}^\mu & \equiv \bigl( \cal{Q}' - \cal{P}(\tau_2 + \delta
	\tau_2) \bigr)^\mu, \\ \delta x^\mu & \equiv \bigl( \cal{Q}' -
	\cal{Q} \bigr)^\mu; \\
\end{split}
\end{equation}
both ${r_2}^\mu$ and ${r'_2}^\mu$ are null vectors. Since the
displacements are infinitesimal, we can write
\begin{equation}
\bigl( \cal{P}(\tau_2 + \delta \tau_2) - \cal{P}(\tau_2) \bigr)^\mu
= \delta \tau_2 \, u^\mu(\tau_2)
\end{equation}
($u^\mu$ is Axel's four-velocity). Then we have
\begin{equation}
\begin{split}
0 & = |{r'_2}^\mu|^2 =
	|{r_2}^\mu + \delta x^\mu - \delta \tau_2 \, u^\mu|^2 = \\
& = |{r_2}^\mu|^2 + 2 \, {r_2}^\mu (\delta x_\mu - \delta \tau_2 \, u_\mu) + O(\delta \tau^2) = \\
& = 2 {r_2}^\mu (\delta x_\mu - \delta \tau_2 \, u_\mu) + O(\delta \tau^2),
\end{split}
\end{equation}
and
\begin{equation}
\frac{\partial \tau_2}{\partial x^\mu} = \frac{{r_2}_\mu}{{r_2}^\nu u_\nu(\tau_2)}.
\end{equation}
The same relation holds for $\partial \tau_1 / \partial x^\mu$:
\begin{equation}
\frac{\partial \tau_1}{\partial x^\mu} = \frac{{r_1}_\mu}{{r_1}^\nu u_\nu(\tau_1)},
\end{equation}
where ${r_1}^\mu \equiv ( \cal{Q} - \cal{P}(\tau_1) )^\mu$. So we can
write the normal vector to the constant-$\btau$ surface as
\begin{equation}
\frac{\partial \btau}{\partial x^\mu} =
\frac{1}{2} \biggl( 
\frac{\partial \tau_1}{\partial x^\mu} +
\frac{\partial \tau_2}{\partial x^\mu}
\biggr) =
\frac{1}{2} \biggl( 
\frac{{r_1}_\mu}{{r_1}^\nu u_\nu(\tau_1)} +
\frac{{r_2}_\mu}{{r_2}^\nu u_\nu(\tau_2)}
\biggr).
\end{equation}
Furthermore,
\begin{equation}
\biggl| \frac{\partial \btau}{\partial x^\mu} \biggr|^2 = \frac{{r_1}^\mu {r_2}_\mu}{
\bigl( {r_1}^\nu u_\nu(\tau_1) \bigr) \,
\bigl( {r_2}^\nu u_\nu(\tau_2) \bigr) }.
\end{equation}
Looking at Fig.\ \ref{fig:marzke}c, you can convince yourself that
${r_1}^\mu {r_2}_\mu > 0$, ${r_1}^\nu u_\nu(\tau_1) > 0$, and
${r_2}^\nu u_\nu(\tau_2) < 0$ (throughout the paper we set $c=1$ and
take a time-like metric). Consequently, the surfaces of
constant-$\btau$ have normal vectors that are time-like everywhere.
Under appropriate hypotheses of smoothness for the world-line
$\cal{P}(\tau)$, the constant-$\btau$ surfaces will also be
differentiable; altogether, they qualify as space-like.

Whereas the constant-time surfaces obtained by the extended-tetrad
procedure (described in Sec.~\ref{sec:definition}) are always
three-dimensional planes, \emph{the global shape of the Märzke-Wheeler
constant-$\btau$ surfaces depends on the entire history of the
observer, both past and future}. Accordingly, the three-dimensional
metric $\sigmag_{ij}$ induced by the Minkowski metric on the surfaces
will depend on $\btau$. This is true in general, but not for
\emph{stationary world-lines}, defined by
\begin{equation}
\label{eq:stationary}
\forall \tau, \; |\cal{P}(\tau + \Delta \tau) - \cal{P}(\tau)| =
|\cal{P}(\tau) - \cal{P}(0)|,
\end{equation}
Stationary world-lines represent motions that show the same behavior
at all proper times; in this case, the surfaces $\Sigma_\btau$ always
maintain the same shape and metric. Synge \cite{Synge67} and Letaw
\cite{Letaw81} obtained stationary world-lines by the alternative
definition of relativistic trajectories with constant acceleration and
curvatures.  In App.~\ref{app:synge}, we briefly review their
classification, as given by Synge \cite{Synge67}.

You can easily build a stationary trajectory by taking any time-like
integral curve of the isometries of Minkowski space-time, and
rescaling its parametrization to obtain a world-line that satisfies
$u^\mu u_\mu = -1$. Indeed, in this way we can obtain \emph{any}
stationary trajectory, because we can always write its four-velocity
as a linear combination $U^\mu$ of the ten Minkowski Killing
fields\footnote{They are the four translations $\partial_t$,
$\partial_x$, $\partial_y$, $\partial_z$, the three boosts $x \,
\partial_t + t \, \partial_x$, $y \, \partial_t + t \, \partial_y$, $z
\, \partial_t + t \, \partial_z$, and the three rotations, $y \,
\partial_z - z \, \partial_y$, $z \, \partial_x - x \, \partial_z$ $x
\, \partial_y - y \, \partial_x$.} (i.~e., the infinitesimal
generators of isometries).  The simplest case of stationary
trajectories are inertial world-lines, obtained by combining the
Killing fields of a time translation and a space translation; further
examples are linear uniform acceleration and uniform rotation,
obtained as the integral curves of, respectively, a Lorentz boost and
a rotation plus time translation.

No matter how we choose to define the constant-time surfaces of a
stationary observer (call her ``Stacy''), the Killing field $U^\mu$
(which coincides with $u^\mu$ on Stacy's world-line, but is defined
all over Minkowski space-time) generates infinitesimal translations in
time that carry each constant-time surface into the next one, while
conserving its three-metric. Once Stacy has chosen a \emph{single}
constant-time surface and a set of spatial coordinates to describe it,
she can use $U^\mu$ to propagate the surface and its coordinates
forward and backward in time, defining coordinates for the entire
Minkowski space-time.

\section{Märzke-Wheeler coordinates for stationary observers: examples}

\noindent Stationary curves are a very useful arena to compare
Märzke-Wheeler coordinates with other accelerated systems, such as the
stationary coordinates derived by Letaw and Pfautsch
\cite{Letaw82}. As a first example, suppose Stacy moves with linear,
uniform acceleration in (1+1)-dimensional Minkowski
space-time\footnote{Also known as \emph{hyperbolic motion}; in Synge's
classification \cite{Synge67}, a \emph{type IIa helix}.}. We can write
the trajectory as
\begin{equation}
\left\{
\begin{aligned}
t&=g^{-1}\sinh g\tau, \\
x&=g^{-1}\cosh g\tau,
\end{aligned}
\right.
\quad \text{(\emph{Hyper-Stacy}: world-line)}
\end{equation}
which is an integral curve of the infinitesimal Lorentz boost $U^\mu =
g (x \, \partial_t + t \, \partial_x)$, where $g$ is the magnitude of
the acceleration. In this case, the extended-tetrad procedure gives
the traditional Rindler coordinates \cite{Rindler75}:
\begin{equation}
\left\{
\begin{aligned}
\mbox{} & t &&= g^{-1} (1+\xi) \sinh g \tau, \\
&x && = g^{-1} (1+\xi) \cosh g \tau.
\end{aligned}
\right.
\quad \text{(\emph{Hyper-Stacy}: Rindler coordinates)}
\end{equation}
You can check easily that the flow of $U^\mu$ carries the
constant-$\tau$ surfaces backward and forward in $\tau$, and that the
\emph{Rindler metric} $ds^2 = -(1 + g \xi)^2 d\tau^2 + d\xi^2$ is
always conserved. Let us now derive Märzke-Wheeler coordinates for
Hyper-Stacy's motion. According to our prescriptions, the surface
$\Sigma_{\btau=0}$ [the set of the events that are simultaneous to
${\cal P}(0)$] includes all the events that, for some $\sigma$,
receive light signals from ${\cal P}(-\sigma)$ and send them back to
${\cal P}(\sigma)$. By symmetry, $\Sigma_{\btau=0}$ must coincide with
the positive-$x$ semiaxis; we then find that the Märzke-Wheeler radial
coordinate is $\sigma=g^{-1} \log g x$. Using the finite isometry
generated by $U^\mu$ with parameter $\btau'$, we can now turn
$\Sigma_{\btau = 0}$ into any other $\Sigma_{\btau'}$. Altogether, the
coordinate transformation between Minkowski and Märzke-Wheeler
coordinates is
\begin{equation}
\left\{
\begin{aligned}
\mbox{} & t &&= g^{-1} e^{g \sigma} \sinh g \bar{\tau}, \\
&x && = g^{-1} e^{g \sigma} \cosh g \bar{\tau}.
\end{aligned}
\right.
\quad \text{(\emph{Hyper-Stacy}: M.-W.\ coordinates)}
\end{equation}
The Rindler and Märzke-Wheeler constant-time surfaces coincide, and
indeed the two coordinate sets are very similar. (If we identify $\xi$
with $\sigma$, and $\tau$ with $\btau$, they coincide up to linear
order, because both systems must coincide with local Lorentz frames in
the vicinity of the world-line).

We turn now to a more interesting example, where Märzke-Wheeler
coordinates diverge from conventional wisdom: uniform relativistic
rotation\footnote{In Synge's classification \cite{Synge67}, a
\emph{type IIc helix}.}. A typical trajectory in 2+1 dimensions for
``Roto-Stacy'' is
\begin{equation}
\label{eq:rsworldline}
\left\{
\begin{aligned}
\mbox{} & t &&= \sqrt{1 + R^2 \Omega^2} \, \tau, \\
&r && = R, \\
&\phi && = \Omega \, \tau,
\end{aligned}
\right.
\quad \text{(\emph{Roto-Stacy}: world-line)}
\end{equation}
where the constant $R$ is the geometric radius of the trajectory, and
$\Omega$ is the proper angular velocity; the coordinate angular
velocity is $d\phi/dt = \Omega / \sqrt{1 + \Omega^2
R^2}$. Finally, Roto-Stacy's generating Killing vector field is $U^\mu =
\sqrt{1 + R^2 \Omega^2} \, \partial_t + \Omega \, \partial_\phi$. The
traditional coordinate system for Roto-Stacy are rigidly rotating
coordinates:
\begin{equation}
\label{eq:rigidly}
\left\{
\begin{aligned}
\mbox{} & t &&= \sqrt{1 + R^2 \Omega^2} \, \tau, \\
&r && = r', \\
&\phi && = \phi' + \Omega \, \tau
\end{aligned}
\right.
\quad \text{(\emph{Roto-Stacy}: rigidly rotating coordinates)}
\end{equation}
(some authors even define $t=\tau$, violating the first requirement we
set in Sec.~\ref{sec:definition}). In these coordinates, Roto-Stacy
stands fixed in space at $r' = R$, $\phi' = 0$; the constant-$\tau$
surfaces coincide with constant-$t$ planes; and the points with fixed
$r'$ and $\phi'$ rotate in the inertial frame with angular velocity
$d\phi/dt = \Omega / \sqrt{1 + \Omega^2 R^2}$, which is faster than
light for $r' > r'_{\mathrm{lim}} = \sqrt{1 + \Omega^2 R^2} /
\Omega^2$. The metric is
\begin{equation}
\begin{gathered}
\begin{aligned}
\mbox{} & ds^2 &=& -(1 + \Omega^2 R^2) \, d\tau^2 + {r'}^2 (d \phi' + \Omega \, d\tau)^2 + d{r'}^2 = \\
& &=& -[1 + (R^2 - {r'}^2) \, \Omega^2] \, d\tau^2 + 2 \, \Omega \, {r'}^2 \, d\tau \, d\phi'+ {r'}^2 d{\phi'}^2 + d{r'}^2
\end{aligned} \\
\text{(\emph{Roto-Stacy}: rigidly rotating metric)}
\end{gathered}
\end{equation}
\begin{figure}
\includegraphics{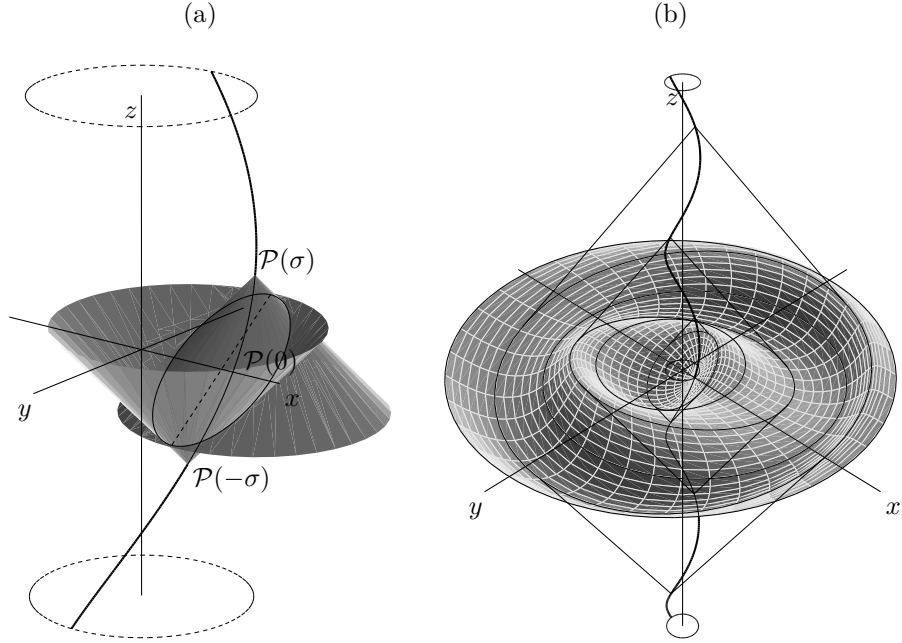}
\caption{Geometry of constant-$\btau$ surfaces for uniformly rotating
observers. (a): Intersection of the light cones with origin in
$\cal{P}(-\sigma)$ and $\cal{P}(\sigma)$ defines an ellipse. (b) Union
of all constant-$\sigma$ ellipses weaves the constant-$\btau$
surface. Notice the oscillating pitch of the
ellipses.\label{fig:rotosurface}}
\end{figure}

Now move on to Märzke-Wheeler coordinates, and consider first the
surface $\Sigma_{\btau = 0}$. Märzke-Wheeler coordinates have their
origin at Roto-Stacy's position, $\cal{P}(0)$: $(x = R, \, y = 0)$.
We find the curves of constant $\sigma$ as the intersection (an
ellipse) of the future light cone of $\cal{P}(-\sigma)$ with the past
light cone of $\cal{P}(\sigma)$. As $\sigma$ increases, the ellipses
move outward, weaving the surface $\Sigma_{\btau = 0}$, which turns
out to be defined by (see App.~\ref{app:typeiic}):
\begin{equation}
\label{eq:rszero}
\left\{
\begin{aligned}
\mbox{} & t &&= c(\sigma) \sin \theta, \\
&x && = b(\sigma) \cos \theta + R \cos \Omega \sigma, \\
&y && = a(\sigma) \sin \theta,
\end{aligned}
\right. \; \text{(\emph{Roto-Stacy}: M.-W.\ coord., $\btau=0$)}
\end{equation}
where $a(\sigma) = \sqrt{1 + R^2 \Omega^2} \, \sigma$, $c(\sigma) = R
\sin \Omega \sigma$, and $b(\sigma) = \sqrt{a^2(\sigma) -
c^2(\sigma)}$ (our choice of the angular coordinate is conventional,
but convenient). As $\sigma$ increases, the centers of the ellipses
oscillate on the $x$-axis between $R$ and $-R$; the semi-axes
$a(\sigma)$ and $b(\sigma)$ grow in such a way that no two ellipses
ever intersect; and the ellipses themselves pitch up and down in the
time direction, as if they were hinging on the $y$-axis (see Fig.\
\ref{fig:rotosurface}), so the Märzke-Wheeler constant-$\btau$ surface
$\Sigma_{\btau = 0}$ deviates in undulatory fashion with respect to
the Minkowski constant-time surface $t = 0$ [because any event
$\cal{Q}$ looks closer if the emission and detection events,
$\cal{P}(-\sigma)$ and $\cal{P}(\sigma)$, are on the near side of the
origin; it looks farther if they are on the other side]. In the limit
$\sigma \rightarrow \infty$, the constant-$\sigma$ ellipses turn into
circles; but the undulation in the $t$ direction maintains the finite
amplitude $R$.

We use the isometry generated by $U^\mu$ to propagate these
coordinates from $\Sigma_{\btau = 0}$ throughout Minkowski
space-time. The complete transformation between Minkowski and
Märzke-Wheeler coordinates is then
\begin{equation}
\label{eq:rscomplete}
\left\{
\begin{aligned}
\mbox{} &
\begin{array}{c}
t
\end{array} = \;
c(\sigma) \sin \theta + \sqrt{1 + R^2 \Omega^2} \, 
\tau, \\
& \left( \begin{array}{c}
	x \\ y
\end{array} \right) = 
\left( \begin{array}{cc}
	\cos \Omega \tau & -\sin \Omega \tau \\
	\sin \Omega \tau & \cos \Omega \tau
\end{array} \right) \cdot
\left( \begin{array}{c}
  b(\sigma) \cos \theta + R \cos \Omega \sigma \\
	a(\sigma) \sin \theta
\end{array} \right).
\end{aligned}
\right.
\end{equation}
\section{Märzke-Wheeler coordinates and the relativistic paradox of the twins}

\begin{figure}
\includegraphics{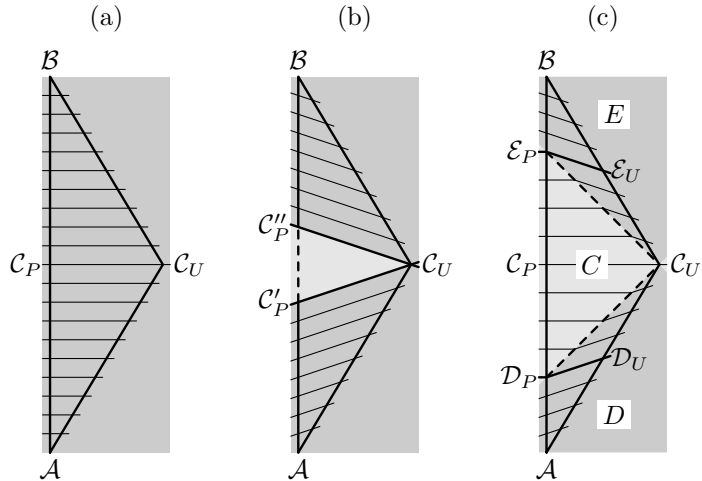}
\caption{The relativistic {\em paradox of the twins}.  The
world-lines of the twins are drawn in the Lorentz rest frame of the
inertial twin, Penelope, who moves in space-time from $\cal A$ to
$\cal B$ through ${\cal C}_P$. The journeying twin, Ulysses, travels
first from $\cal A$ to ${\cal C}_U$ with velocity $v$, then inverts
his motion to rejoin Penelope in $\cal B$.  (a): Lorentz slicing of
space-time according to Penelope.  (b): Lorentz slicing of space-time
according to Ulysses, on his separate stretches of inertial
motion. Ulysses skips a finite lapse of Penelope's world-line (shown
dashed).  (c): Märzke-Wheeler slicing of space-time, according to
Ulysses. This slicing coincides with the Lorentz slicing in (b) for
events in the regions $D$ and $E$ (these events belong to the causal
envelopes of the world-line segments $\cal{A}\cal{C}_U$ and
$\cal{C}\cal{B}_U$), but it shows a peculiar structure in region $C$.
\label{fig:twins}}
\end{figure}

\noindent Märzke-Wheeler coordinates cast a new light on the
relativistic \emph{paradox of the twins}\footnote{The literature on
the subject is immense and often redundant. Even if the paradox was
already present in Einstein's 1905 seminal paper \cite{Einstein05}, it
was P.\ Langevin who first presented it in its modern form. Arzeliès
\cite{Arzelies66} and Marder \cite{Marder71} give excellent annotated
bibliographies for contributions up to, respectively, 1966 and 1971.}.
This \emph{gedankenexperiment} earns the designation of ``paradox''
because, at first sight, the motion of the twins is reciprocal,
whereas the physical effects of relativistic time dilation are not.
In the usual arrangement, shown in Fig.~\ref{fig:twins}a, the
journeying twin (``Ulysses'') moves with constant speed $v$, first
away from and then toward the waiting, inertial (and non-identical!)
twin ``Penelope''.  According to the Lorentz transformation between
the inertial frames associated with the twins, Penelope sees Ulysses'
proper time as dilated by the relativistic factor $\gamma = (1 -
v^2)^{-1/2} > 1$, so when the twins are rejoined, Penelope has aged
$\gamma^{-1}$ times as much as Ulysses.  Yet, it is also true that
Penelope always moves with a speed $v$ relatively to Ulysses, so
\emph{he} should see \emph{her} proper time as dilated!

The problem is that the notion of time dilation, as it is usually
discussed, amounts to little more than a statement on how to relate
the coordinate times of different Lorentz frames; it also concerns the
observations of different inertial observers, whose proper times
coincide with the coordinate times of their Lorentz rest frames.  Now,
Ulysses is \emph{not} an inertial observer \emph{throughout} his
motion, because at event $\cal{C}_U$ he turns around and begins his
return trip towards Penelope. Along the world-line segments
$\cal{A}\cal{C}_U$ and $\cal{C}\cal{B}_U$, it is correct to say that
Ulysses sees Penelope's proper time as dilated, in the following
sense: if Ulysses compares his proper time with Penelope's at events
which are simultaneous in his Lorentz rest frame, then Penelope
appears to be aging at a slower pace.  However, when Ulysses inverts
his velocity at $\cal{C}_U$ (see Fig.~\ref{fig:twins}b), he switches
to a new Lorentz frame, and his constant-time surfaces change their
space-time orientation abruptly.  Just before arriving in $\cal{C}_U$,
Ulysses considers himself simultaneous to the event $\cal{C}'_P$ along
Penelope's world-line; just after leaving $\cal{C}_U$, according to
his new Lorentz frame, Ulysses considers himself simultaneous to
$\cal{C}''_P$. However, $\cal{C}'_P$ and $\cal{C}''_P$ are distinct
events, separated by a finite lapse of time! \emph{There is a finite
section of Penelope's world-line which Ulysses effectively skips and
to which he is never simultaneous}. Because of this missing finite
lapse of Penelope's proper time, Ulysses is younger at his final
reunion with Penelope, even if throughout the trip he reckoned that
Penelope was aging at a slower pace than him\footnote{Ulysses'
``switch'' of Lorentz frames in $\cal{C}_U$ has generated some
controversy, centered on the physical effects of Ulysses' acceleration
around $\cal{C}_U$. These effects are irrelevant, as can be seen by
the ``third twin'' argument introduced by Lord Halsbury
\cite{Salmon75}: in brief, at $\cal{C}_U$ Ulysses communicates the
reading of his clock to a third twin who was already traveling towards
Penelope with velocity $v$; thus, the proper time elapsed on the
different paths $\cal{A}\cal{C}_U\cal{B}$ and
$\cal{A}\cal{C}_P\cal{B}$ can be compared without any twin ever
experiencing acceleration.}!

From a general-relativistic perspective, there is no paradox from the
beginning: Ulysses and Penelope move on different space-time paths
between the same two events. The lapse of proper time is a particular
functional of the path followed: no wonder that it is different for
the two twins! The surprise of non-reciprocal time dilation arises
because (a) Ulysses needs to compare \emph{simultaneous} events on his
and on Penelope's world-lines to know who is aging faster, so he needs
a global notion of simultaneity or, equivalently, a slicing of
space-time into space-like, constant-time surfaces; (b) since our
Ulysses has a special-relativistic background, he naturally employs
the slicing implicit in his two distinct Lorentz rest frames; (c) but
that slicing fails to cover a finite region of space-time, where
nevertheless Penelope spends part of her time!

Märzke-Wheeler coordinates avoid this problem, since by definition
they provide a consistent time slicing of the causal envelope of any
observer's world-line: Ulysses and Penelope stay well inside each
other's causal envelope, simply because they start together and cannot
travel faster than light. For inertial Penelope, Märzke-Wheeler
coordinates reproduce a Lorentz rest frame (Fig.~\ref{fig:twins}a). So
nothing changes in her account of Ulysses' aging: her proper time
lapse $\Delta t_P$ is $\gamma^{-1}$ times Ulysses' proper time lapse
$\Delta t_U$.

Likewise, Märzke-Wheeler coordinates for Ulysses do reproduce a
Lorentz frame, but only and separately for the events in the causal
envelopes ($D$ and $E$) of the segments of Ulysses' uninterrupted
inertial motion ($\cal{A}\cal{C}$ and $\cal{C}\cal{B}$; see
Fig.~\ref{fig:twins}c). In the process of Märzke-Wheeler
synchronization, the events in $D$ and $E$ communicate with events
along the \emph{same} segment.  On the contrary, region $C$ contains
events that are space-like related to $\cal{C}$, and that receive
light signals from $\cal{A}\cal{C}$ and reflect them back to
$\cal{C}\cal{B}$. It is in this region that the non-inertial character
of Ulysses' motion becomes manifest. A simple calculation
(App.~\ref{app:twins}) yields the slicing structure shown in
Fig.~\ref{fig:twins}c: in $D$ and $E$ the slices assume the typical
inclination of Lorentz constant-time surfaces, but in $C$ the slices
become perpendicular to $\cal{A}\cal{B}$ (Penelope's world-line),
because they split the difference between the two opposing inertial
motions $\cal{A}\cal{C}$ and $\cal{C}\cal{B}$.

\begin{figure}
\includegraphics{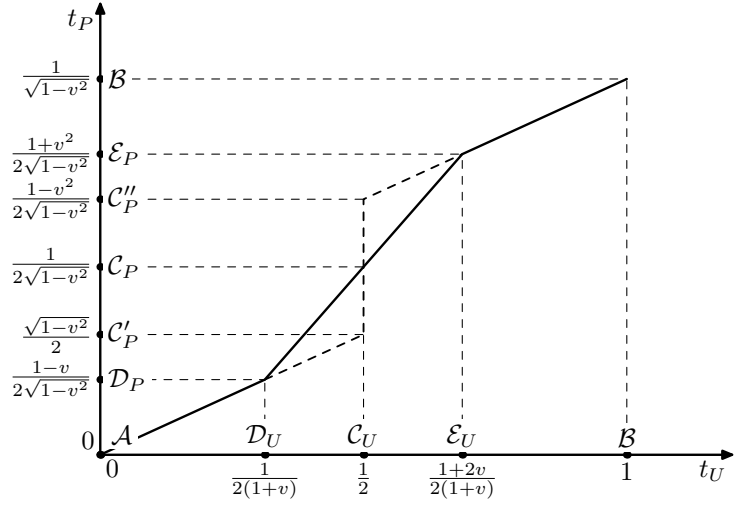}
\caption{Penelope's proper-time, in units of Ulysses' total
proper-time lapse, as determined by Ulysses through Lorentz slicing
(dashed line, see Fig.~\protect\ref{fig:twins}b) or Märzke-Wheeler
slicing (continuous line, see
Fig.~\protect\ref{fig:twins}c).\label{fig:lapses}}
\end{figure}

If Ulysses employs the Märzke-Wheeler notion of simultaneity to
compare his age with Penelope's at simultaneous times, he accounts for
the final aging difference as follows. As long as Penelope's
trajectory remains within the regions $D$ and $E$ where the
Märzke-Wheeler and Lorentz notions of simultaneity coincide, Ulysses
ages $\gamma$ times faster than Penelope, just as a naïve use of
relativistic time dilation would imply. However, when Penelope moves
through region $C$ (from $\cal{D}_P$ to $\cal{E}_P$), she ages
$\gamma^{-1} (1-v)^{-1} > 1$ times faster than Ulysses (who moves from
$\cal{D}_U$ to $\cal{E}_U$). Altogether, when the twins are rejoined
in $\cal{B}$, Ulysses is younger by an overall factor of $\gamma$. See
Table~\ref{tab:twins} and Fig.~\ref{fig:lapses} for a precise tally of
proper times. In App.\ \ref{app:twinsb} we study Ulysses'
Märzke-Wheeler interpretation of Penelope's aging in a modified
construction where Ulysses moves with constant speed and acceleration
on \emph{Roto-Stacy}'s circular trajectory. The resulting $t_P[t_U]$
(Fig.\ \ref{fig:twinrot}) is smooth and resembles qualitatively the
function shown in Fig.\ \ref{fig:lapses}.
\begin{table}
\centering
\begin{tabular}{l||c|c|c|c|c}
& \parbox{0.18\textwidth}{\small \centering Ulysses' $\Delta t_U$ \\ in segment}
& \parbox{0.11\textwidth}{\small \centering Ulysses' total $t_U$}
& \parbox{0.21\textwidth}{\small \centering Penelope's $\Delta t_P$ \\ in segment}
& \parbox{0.13\textwidth}{\small \centering Penelope's total $t_P$} 
& \parbox{0.15\textwidth}{\small \centering $\frac{d t_P}{d t_U}$ \\ in segment} \\
\hline \hline
$\cal{A}\cal{D}$ &
$\frac{1}{2(1+v)}$ &
$\frac{1}{2(1+v)}$ &
$\frac{1}{2} \frac{1-v}{\sqrt{1-v^2}}$ &
$\frac{1}{2} \frac{1-v}{\sqrt{1-v^2}}$ &
$\sqrt{1-v^2}$ \\
\hline
$\cal{D}\cal{C}$ &
$\frac{v}{2(1+v)}$ &
$\frac{1}{2}$ &
$\frac{1}{2} \frac{v}{\sqrt{1-v^2}}$ &
$\frac{1}{2} \frac{1}{\sqrt{1-v^2}}$ &
$\frac{1+v}{\sqrt{1-v^2}}$ \\
\hline
$\cal{C}\cal{E}$ &
$\frac{v}{2(1+v)}$ &
$\frac{1+2v}{2 (1+v)}$ &
$\frac{1}{2} \frac{v}{\sqrt{1-v^2}}$ &
$\frac{1}{2} \frac{1+v^2}{\sqrt{1-v^2}}$ &
$\frac{1+v}{\sqrt{1-v^2}}$ \\
\hline
$\cal{E}\cal{B}$ &
$\frac{1}{2(1+v)}$ &
$1$ &
$\frac{1}{2} \frac{1-v}{\sqrt{1-v^2}}$ &
$\frac{1}{\sqrt{1-v^2}}$ &
$\sqrt{1-v^2}$ \\
\end{tabular}
\vspace{12pt}
\caption{Evolution of Ulysses' and Penelope's proper times along the
segments shown in Fig.~\protect\ref{fig:twins}c; all comparisons are
made at events that are simultaneous according to Ulysses'
Märzke-Wheeler slicing. The last column shows that Ulysses' ages
faster than Penelope's along segments $\cal{A}\cal{D}$ and
$\cal{E}\cal{B}$, but not along $\cal{D}\cal{C}$ and
$\cal{C}\cal{E}$. Units are normalized so that Ulysses' total proper
time lapse is $1$.\label{tab:twins}}
\end{table}

Keep in mind that the comparison of local relative aging is dependent
on how we slice space-time into constant-time surfaces. Alternative
slicings will lead Ulysses to \emph{different distributions} of
Penelope's total proper time along his world-line.  Stautberg
Greenwood \cite{Greenwood76} defines simultaneity by integrating the
Doppler-shifted frequency of monocromatic signals exchanged by the
twins. Unruh \cite{Unruh81} employs the notion of \emph{parallax
distance} to extend Ulysses' local definitions of space and time, to
the effect that at times he sees Penelope recede in time.  Debs and
Redhead \cite{Debs96} analyze the class of slicings induced by
Reichenbach's \emph{non-standard synchronies} \cite{Redhead93}, which
generalize the Einstein convention by positing different speeds for
the light signals in the two directions. However, we believe that
Märzke-Wheeler slicing has a simple physical rationale and that it
does a good job of locating the non-reciprocal, differential aging in
the region of space-time where the non-local effects of Ulysses'
turn-around in $\cal{C}$ are felt.

\section{Conclusions}

\noindent We have shown how to use the Märzke-Wheeler construction to
build accelerated systems of coordinates that are \emph{adapted to the
motion of an arbitrary observer in flat space-time}, in the sense
that: (a) on the observer's world-line, the Märzke-Wheeler time
coordinate $\btau$ coincides with the observer's proper time $\tau$;
(b) in a neighborhood of the world-line, Märzke-Wheeler coordinates
reduce to Lorentz (spherical) coordinates; (c) the procedure assigns
smoothly and unambiguously a time $\btau$ and a Märzke-Wheeler radial
coordinate $\sigma$ to all events in the \emph{causal envelope} of the
world-line (that is, to all events from which bi-directional
communication with the observer is possible). This is obtained with a
simple geometric construction that generalizes the Einstein
synchronization criterion. In particular, we showed that $\btau$
indexes a smooth foliation of the causal envelope of the world-line
into space-like surfaces.

The Märzke-Wheeler construction is intrinsically global: \emph{the
structure of any constant-$\btau$ surface depends on the geometry of
the entire world-line of the observer}. Yet this global dependence is
hierarchical. Take for instance the constant-time surface $\btau =
\tau_0$, with origin in $\cal{P}(\tau_0)$: the behavior of the
world-line at proper times that lie to the future of $\tau_0 + \Delta
\tau$, or to the past of $\tau_0 - \Delta \tau$, can only influence
the structure of the constant-time surface for $\sigma > \Delta \tau$.

We have examined the special case of stationary observers, where
Märzke-Wheeler constant-time surfaces are all identical, and they are
translated into each other by a family of Minkowski space-time
isometries. In the simplest case, hyperbolic motion, Märzke-Wheeler
coordinates are related to the familiar Rindler system by a monotonic
map between the radial coordinates. In the case of circular motion,
however, the Märzke-Wheeler constant-$\btau$ surfaces have a much
richer structure than the constant-time planes of rigidly rotating
coordinates.

Finally, we have discussed how to use the notion of
\emph{Märzke-Wheeler simultaneity} to elucidate the relativistic
paradox of the twins, by establishing a continuous correspondence
between the lapses of proper time experienced by the twins. It is
possible to attribute the differential aging of the twins to distinct
segments of their world-lines, where we can conclude that one twin is
aging faster. Although this attribution is not unique, it is justified
physically by recourse to generalized Einstein synchronization, and \emph{it is not possible with other definitions of simultaneity} (such as a naïve use of instantaneous Lorentz frames).

\begin{appendix}

\section{Stationary trajectories in flat space-time (Synge's helixes)}
\label{app:synge}
\begin{figure}
\includegraphics{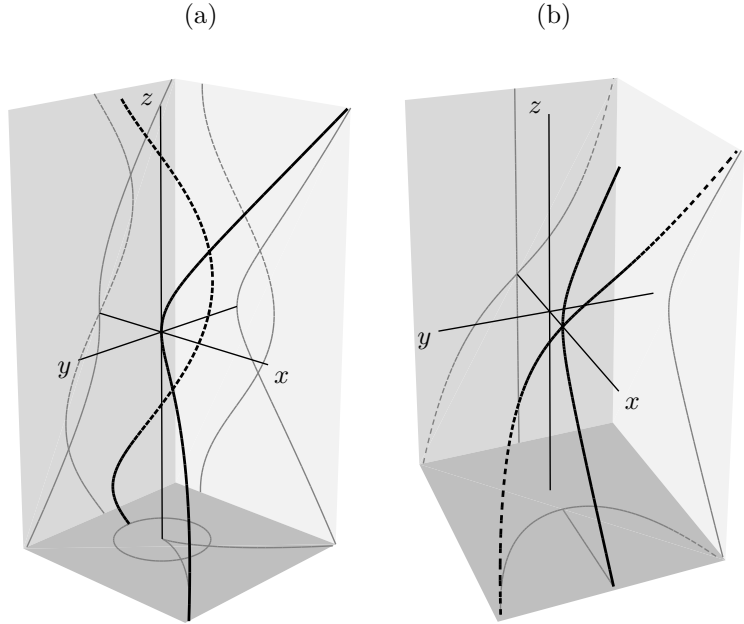}
\caption{Synge's helixes. (a): Type IIc (shown dashed), and type IIb
helixes. (b): Type IIa (dashed), and type III helixes. Notice the cusp
in the $xy$-plane projection of the type IIb curve; also notice that
the type III helix coincides with projection of the type IIa curve on
the $xz$-plane.\label{fig:synge}}
\end{figure}

\noindent Synge \cite{Synge67} solved the \emph{relativistic
Frenet-Serret equations},
\begin{equation}
\left\{
\begin{aligned}
\mbox{} &\dot{u}^\mu && =c_1 n_1^\mu, \\
&\dot{n}_1^\mu && = c_2 n_2^\mu + c_1 u^\mu, \\
&\dot{n}_2^\mu && = c_3 n_3^\mu - c_2 n_1^\mu, \\
&\dot{n}_3^\mu && = - c_3 n_2^\mu
\end{aligned}
\right.
\label{eq:frenet}
\end{equation}
(where $u^\mu$ is the 4-velocity and $n_i^\mu$ are the three
\emph{normals}), by restricting the curvature coefficients $c_1$,
$c_2$, and $c_3$ to constants. We briefly summarize Synge's
classification of the resulting trajectories (for pedagogical
purposes, we invert Synge's enumeration). In Fig.~\ref{fig:synge}, we
show examples of these curves.
\subsection*{Inertial world-lines (type IV)} All curvatures vanish.
\subsection*{Hyperbolic motion (type III)} (\emph{Hyper-Stacy}) $c_2 =
c_3 = 0$. The only non-zero curvature is the acceleration. Motion is
restricted to a $(1+1)$-dimensional hyperplane; the trajectory is
spatially unlimited and the 3-velocity approaches the speed of light
asymptotically. In a suitable Lorentz frame, we can write the
world-line as
\begin{equation}
\left\{
\begin{aligned}
t&=c_1^{-1}\sinh c_1\tau, \\
x&=c_1^{-1}\cosh c_1\tau, \\
y&=z=0,
\end{aligned}
\right.
\quad \text{(Type III)}
\end{equation}
where $\tau$ is proper time, and $c_1$ is the magnitude of the acceleration.
\subsection*{Plane helixes (type II)} Only $c_3 = 0$: the spatial
curvature $c_2$ allows non-trivial motion in a $(2+1)$-dimensional
hyper-plane. There are three subtypes.
	\subsubsection*{Uniform circular motion (Roto-Stacy, type IIc)}
	If $c_2^2 - c_1^2 > 0$, the world-line winds up in a spatially
	limited domain. It is a circular helix of radius $c_1 / (c_2^2
	- c_1^2)$ and angular velocity $\sqrt{c_2^2 - c_1^2}$.
	\begin{equation}
	\left\{
	\begin{aligned}
	t&=\frac{c_2}{c_2^2 - c_1^2} \, \sqrt{c_2^2 - c_1^2} \, \tau, \\
	x&=\frac{c_1}{c_2^2 - c_1^2} \, \cos \sqrt{c_2^2 - c_1^2} \, \tau, \\
	y&=\frac{c_1}{c_2^2 - c_1^2} \, \sin \sqrt{c_2^2 - c_1^2} \, \tau, \\
	z&=0.
	\end{aligned}
	\right.
	\quad \text{(Type IIc)}
	\end{equation}
	\subsubsection*{Cusped motion (type IIb)} If $c_1^2 - c_2^2 =
	0$, the result is a run-away curve (although it approaches
	spatial infinity only cubically in time and not exponentially
	as type III does), with a peculiar cusp.
	\begin{equation}
	\left\{
	\begin{aligned}
	t&=\tau + \frac{1}{6} \, c_1^2 \tau^3, \\
	x&=\frac{1}{2} \, c_1 \tau^2, \\
	y&=\frac{1}{6} \, c_1^2 \tau^3 \\
	z&=0.
	\end{aligned}
	\right.
	\quad \text{(Type IIb)}
	\end{equation}
	\subsubsection*{Skewed hyperbolic motion (type IIa)} If $c_1^2
	- c_2^2 > 0$, the spatial curvature $c_2$ is not strong enough
	to wind up the world-line, which becomes spatially unlimited
	and approaches asymptotically the speed of light. In fact,
	this solution may be considered as a type III helix combined
	with a linear, uniform motion.
	\begin{equation}
	\left\{
	\begin{aligned}
	t&=\frac{c_1}{c_1^2-c_2^2} \sinh \sqrt{c_1^2-c_2^2} \, \tau, \\
	x&=\frac{c_1}{c_1^2-c_2^2} \cosh \sqrt{c_1^2-c_2^2} \, \tau, \\
	y&=\frac{c_2}{c_1^2-c_2^2} \sqrt{c_1^2-c_2^2} \, \tau, \\
	z&=0.
	\end{aligned}
	\right.
	\quad \text{(Type IIa)}
	\end{equation}
\subsection*{General case (type I)} All curvatures have a finite value,
and the trajectory is truly four-dimensional. The resulting helix is a
product (of sorts) between a type III and a type IIc motion, each of
which takes place in a 2-dimensional hyperplane.
\begin{equation}
\left\{
\begin{aligned}
t & = r \chi^{-1} \sinh \chi \, \tau, \\
x & = q \gamma^{-1} \sin \gamma \, \tau, \\
y & = q \gamma^{-1} \cos \gamma \, \tau, \\
z & = r \chi^{-1} \cosh \chi \, \tau,
\end{aligned}
\right.
\quad \text{(Type I)}
\end{equation}
where
\begin{equation}
\left\{
\begin{aligned}
\chi^2 &= (c_1^2-c_2^2-c_3^2+R)/2, \\
\gamma^2 &= (-c_1^2+c_2^2+c_3^2+R)/2, \\
r^2 &= [(c_1^2+c_2^2+c_3^2)/R + 1]/2, \\
q^2 &= [(c_1^2+c_2^2+c_3^2)/R - 1]/2, \\
R^2 &= (c_1^2-c_2^2-c_3^2)^2 + 4 c_1^2 c_3^2.
\end{aligned}
\right.
\end{equation}

\section{Märzke-Wheeler coordinates for uniformly rotating observers}
\label{app:typeiic}

\begin{figure}
\includegraphics{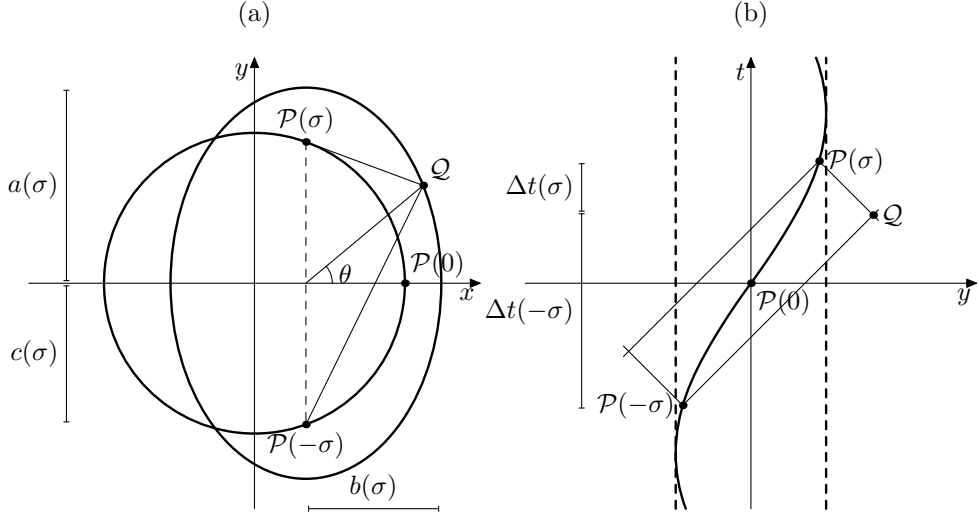}
\caption{Geometric construction of Märzke-Wheeler constant-$\sigma$
surfaces for \emph{Roto-Stacy}. Her world-line's projection is (a) a
circle in the $xy$ plane; (b) a sinusoidal curve in the $yt$ plane
(b).\label{fig:rotocoord}}
\end{figure}

\noindent Roto-Stacy's world-line is given by \eqref{eq:rsworldline},
and in Cartesian coordinates by
\begin{equation}
\label{eq:recartesian}
\left\{
\begin{aligned}
\mbox{} & t &&= \sqrt{1 + R^2 \Omega^2} \, \tau, \\
&x && = R \cos \Omega \tau, \\
&y && = R \sin \Omega \tau.
\end{aligned}
\right.
\quad \text{(\emph{Roto-Stacy}: world-line)}
\end{equation}
We seek equations for the surface $\Sigma_{\btau=0}$, which is
generated by the concentric curves $S(\sigma)$ of constant $\sigma$:
each curve $S(\sigma)$ is defined as the intersection of the future
light cone of ${\cal P}(-\sigma)$ with the past light cone of ${\cal
P}(\sigma)$ (see Fig.~\ref{fig:rotocoord}).

A point ${\cal Q}$ belongs to the future light cone of ${\cal
P}(-\sigma)$ if the spatial distance between ${\cal P}(-\sigma)$ and
${\cal Q}$ equals the coordinate-time difference between them; that
is, if
\begin{equation}
\bigl| \mathbf{x}[\cal{Q}] - \mathbf{x}[\cal{P}(-\sigma)] \bigr| =
t[\cal{Q}] - t[\cal{P}(-\sigma)] = \Delta t(-\sigma);
\end{equation}
a similar relation is true for points on the past light cone of ${\cal
P}(\sigma)$:
\begin{equation}
\bigl| \mathbf{x}[\cal{Q}] - \mathbf{x}[\cal{P}(\sigma)] \bigr| =
t[\cal{P}(\sigma)] - t[\cal{Q}] = \Delta t(\sigma).
\end{equation}
Summing the two equations, we get
\begin{multline}
\bigl| \mathbf{x}[\cal{Q}] - \mathbf{x}[\cal{P}(-\sigma)] \bigr| +
\bigl| \mathbf{x}[\cal{Q}] - \mathbf{x}[\cal{P}(\sigma)] \bigr| = \\
= t[\cal{P}(\sigma)] - t[\cal{P}(-\sigma)]
= \Delta t(\sigma) + \Delta t(-\sigma) = 2 \sqrt{1 + R^2 \Omega^2} \sigma;
\end{multline}
that is, the points on $S(\sigma)$ describe an ellipse in the spatial
plane. These ellipses have ${\cal P}(-\sigma)$ and ${\cal P}(\sigma)$
as their foci, and they are centered in ${\cal C}(\sigma): (x=R \cos
\Omega \sigma, \, y=0)$. We parametrize the ellipses in the obvious
way,
\begin{equation}
\label{eq:rsspatial}
\left\{
\begin{aligned}
\mbox{} & x && = b(\sigma) \cos \theta + R \cos \Omega \sigma, \\
&y && = a(\sigma) \sin \theta.
\end{aligned}
\right.
\quad \text{(ellipses $S(\sigma)$)}
\end{equation}
The length $a(\sigma)$ of the major semiaxis is given by the half-sum
of the distances between any point on $S(\sigma)$ and the two foci:
\begin{equation}
\label{eq:rsa}
a(\sigma) = \frac{1}{2} \Bigl\{
\bigl| \mathbf{x}[\cal{Q}] - \mathbf{x}[\cal{P}(-\sigma)] \bigr| +
\bigl| \mathbf{x}[\cal{Q}] - \mathbf{x}[\cal{P}(\sigma)] \bigr| \Bigr\} =
\sqrt{1 + R^2 \Omega^2} \, \sigma;
\end{equation}
also, from \eqref{eq:recartesian} the half-distance between the foci
is $c(\sigma) = R \sin \Omega \sigma$, so we find the length of the
minor semiaxis $b(\sigma)$ as
\begin{equation}
\label{eq:rsb}
b(\sigma) = \sqrt{a^2(\sigma) - c^2(\sigma)} =
\sqrt{(1 + R^2 \Omega^2) \sigma^2 - R^2 \sin^2 \Omega \sigma}.
\end{equation}
To complete our characterization of $\Sigma_{\btau=0}$, we need only
the coordinate time of the points on the curves $S(\sigma)$. From
Fig.~\ref{fig:rotocoord}b we have
\begin{equation}
\begin{split}
t[\cal{Q}] & =
t[\cal{P}(\sigma)] - \Delta t(\sigma) = 
\frac{1}{2} \bigl\{ \Delta t(-\sigma) + \Delta t(\sigma) \bigr\}
- \Delta t(\sigma) = \\
& = \frac{1}{2} \bigl| \mathbf{x}[\cal{Q}] - \mathbf{x}[\cal{P}(-\sigma)] \bigr| -
\frac{1}{2} \bigl| \mathbf{x}[\cal{Q}] - \mathbf{x}[\cal{P}(\sigma)] \bigr|;
\end{split}
\end{equation}
then by explicit calculation we find that $t[\cal{Q}] = c(\sigma) \sin
\theta$, using \eqref{eq:rsspatial}--\eqref{eq:rsb}. Altogether we
obtain the surface described by \eqref{eq:rszero} and shown in
Fig.~\ref{fig:rotosurface}. Given the symmetry of \emph{Roto-Stacy}'s
motion, the effect of moving from $\Sigma_{\btau=0}$ to
$\Sigma_{\btau=\Delta \btau}$ will be just a translation in $t$ by
$\sqrt{1 + R^2 \Omega^2} \, \Delta \btau$, together with a rotation of
$x$ and $y$ by an angle $\Omega \, \Delta \btau$; the complete
transformation between Märzke-Wheeler and Lorentz coordinates is
therefore that given in \eqref{eq:rscomplete}.

\section{Märzke-Wheeler coordinates for the paradox of the twins: linear motion}
\label{app:twins}

\noindent For simplicity, we use Penelope's Lorentz coordinates to
parametrize Ulysses' world-line (shown in Fig.~\ref{fig:twins}),
putting the origin $(0,0)$ in ${\cal C}_U$, so that the world-line is
described by $x = - v |t|$. Proceeding as in App.~\ref{app:typeiic},
we see that the Märzke-Wheeler constant-time surface that is
simultaneous to $\cal{P}(t_0)$ is given by the events $\cal{Q}$ such
that, for some $s$,
\begin{equation}
\label{eq:calcul}
\left\{
\begin{aligned}
\mbox{} & \bigl| x[\cal{Q}] - x[\cal{P}(t_0 - s)] \bigr| && = 
		t[\cal{Q}] - t[\cal{P}(t_0 - s)], \\
& \bigl| x[\cal{Q}] - x[\cal{P}(t_0 + s)] \bigr| && = 
		t[\cal{P}(t_0 + s)] - t[\cal{Q}].
\end{aligned}
\right.
\end{equation}
We simplify our notation by setting $t = t[\cal{Q}]$ and $x =
x[\cal{Q}]$, and we insert the explicit form of Ulysses' world-line
into \eqref{eq:calcul}:
\begin{equation}
\left\{
\begin{aligned}
\mbox{} & \bigl| x + v |t_0 - s| \bigr| && = 
		t - (t_0 - s), \\
& \bigl| x + v |t_0 + s| \bigr| && = 
		(t_0 + s) - t.
\end{aligned}
\right.
\end{equation}
If we are concerned only with events to the left of Ulysses'
trajectory, the outer absolute values can be exchanged for a
minus. Summing and subtracting the equations, we obtain the following
expressions for $x$ and $t$:
\begin{equation}
\label{eq:simultaneous}
\left\{
\begin{aligned}
- & x && = s && + \frac{1}{2} \bigl( v |t_0 - s| + v |t_0 + s| \bigr), \\
& t && = t_0 && + \frac{1}{2} \bigl( v |t_0 + s| - v |t_0 - s| \bigr).
\end{aligned}
\right.
\quad \text{(events simultaneous to $\cal{P}(t_0)$)}
\end{equation}
Let us take $t_0 > 0$, and examine Eq.~\eqref{eq:simultaneous}: if an
event $\cal{Q}$, simultaneous to $\cal{P}(t_0)$, belongs to region $E$
of Fig.~\ref{fig:twins}c, both $\cal{P}(t_0 - s)$ and $\cal{P}(t_0 +
s)$ will be in region $E$. It follows that $t_0 - s > 0$ and $t_0 + s
> 0$, and therefore
\begin{equation}
\label{eq:rege}
\left\{
\begin{aligned}
- & x && = s && + v t_0, \\
& t && = t_0 && + v s.
\end{aligned}
\right.
\quad \text{(region $E$)}
\end{equation}
In a neighborhood of his world-line, these equations reproduce the
slices of his constant Lorentz time. On the other hand, if $Q$ belongs
to region $C$, then $t_0 - s < 0$, $t_0 + s > 0$, and
\begin{equation}
\label{eq:regc}
\left\{
\begin{aligned}
- & x && = (1 + v) s, \\
& t && = (1 + v) t_0.
\end{aligned}
\right.
\quad \text{(region $C$)}
\end{equation}
These relations create the flat structure of Märzke-Wheeler slices
shown in Fig.~\ref{fig:twins}c. The two coordinate patches of
Eqs.~\eqref{eq:rege}, \eqref{eq:regc} join correctly on $x = -|t|$,
where $s = t_0$.
\section{Märzke-Wheeler coordinates for the paradox of the twins: circular motion}
\label{app:twinsb}
\noindent In this scenario, we make the twins start together at the
event $\cal F$ with Lorentz coordinates $t=0$, $x=R$, and $y=0$. While
the stationary twin Penelope stands fixed in space, Ulysses completes
one circular orbit according to Eqs.\ \eqref{eq:rsworldline} and
\eqref{eq:recartesian}, and rejoins Penelope at the event $\cal G$,
defined by $t = 2 \pi \Omega^{-1} \sqrt{1 + \Omega^2 R^2}$, $x=R$, and
$y=0$. After one revolution, Ulysses' proper time lapse is $\Delta
\tau = 2 \pi \Omega^{-1}$; Penelope's proper time coincides with the
Lorentz coordinate time, so that her proper time lapse is $\sqrt{1 +
\Omega^2 R^2}$ times Ulysses'. It turns out that this coefficient is
just $\gamma = (1 - v^2)^{-1/2}$, because Ulysses moves with a
constant velocity $v = \Omega R / \sqrt{1+\Omega^2 R^2}$. In the end,
we get the same differential aging of the twins as in the simpler
linear geometry of App.\ \ref{app:twins}, and as predicted by a naïve
application of the time dilation rule.

\begin{figure}
\includegraphics{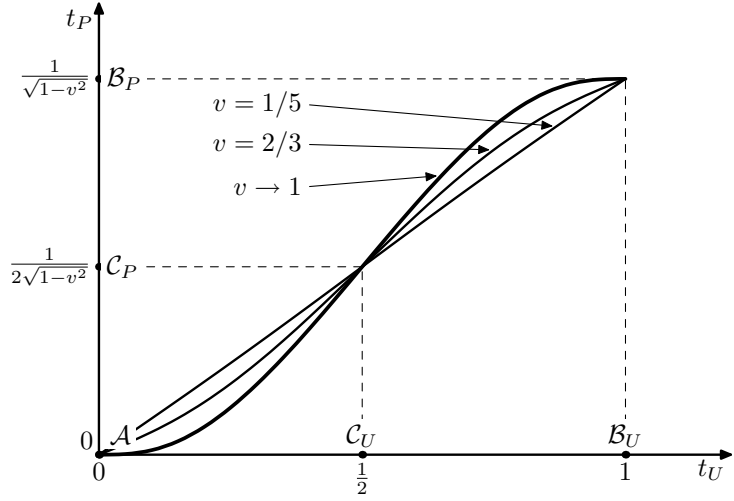}
\caption{Circular version of the paradox of the twins. The graph
shows Penelope's proper-time, in units of Ulysses' total proper-time
lapse, as determined by Ulysses through Märzke-Wheeler slicing. In
these renormalized units, the shape of the curve depends only on
Ulysses' velocity. For $v \rightarrow 0$, the curve tend to a straight
line; for $v \rightarrow 1$, to a limit curve.\label{fig:twinrot}}
\end{figure}

To study the local distribution of this differential aging, we need to
determine the Märzke-Wheeler time (according to Ulysses) of all the
events on Penelope's world-line. It is expedient to work in Lorentz
polar coordinates centered around Penelope's location. Then Ulysses'
world-line is given by
\begin{equation}
\label{eq:urworldline}
\left\{
\begin{aligned}
\mbox{} & t &&= \sqrt{1 + R^2 \Omega^2} \, \tau, \\
&\rho && = 2 R \sin \frac{\Omega \tau}{2}, \\
&\theta && = \frac{\pi}{2} - \frac{\Omega \tau}{2}.
\end{aligned}
\right.
\quad \text{(\emph{Roto-Ulysses}: world-line)}
\end{equation}
Let us now proceed in analogy with App.\ \ref{app:twins}. Eliminating
the parameter $\tau$, we describe Ulysses' world-line as
\begin{equation}
\label{eq:rutraj}
{\cal P}(t): \rho = 2 R \, \sin \biggl( \frac{\Omega t}{2 \sqrt{1 + R^2 \Omega^2}} \biggr).
\quad \text{(\emph{Roto-Ulysses}: world-line)}
\end{equation}
If we take only target events $\cal{Q}$ on Penelope's world-line, the
light-cone conditions \eqref{eq:calcul} can be restated simply as
\begin{equation}
\left\{
\begin{aligned}
\mbox{} & \rho[\cal{P}(t_0 - s)] && = 
		t[\cal{Q}] - t[\cal{P}(t_0 - s)], \\
& \rho[\cal{P}(t_0 + s)] && = 
		t[\cal{P}(t_0 + s)] - t[\cal{Q}],
\end{aligned}
\right.
\end{equation}
where $t_0$ identifies an event along Ulysses' world-line, and $t$ and
$s$ identify the simultaneous event (in the Märzke-Wheeler sense)
along Penelope's world-line. Now, set $t = t[\cal{Q}]$ and use Eq.\
\eqref{eq:rutraj}:
\begin{equation}
\left\{
\begin{aligned}
\mbox{} & 2 R \, \sin \biggl( \frac{\Omega (t_0 - s)}{2 \sqrt{1 + \Omega^2 R^2}} \biggr)  =	t - t_0 + s, \\
& 2 R \, \sin \biggl( \frac{\Omega (t_0 + s)}{2 \sqrt{1 + \Omega^2 R^2}} \biggr) =
		t_0 + s - t.
\end{aligned}
\right.
\end{equation}
We sum and subtract these two equations, and rearrange their terms:
\begin{equation}
\label{eq:finaltwin}
\left\{
\begin{aligned}
\mbox{} & t = t_0 -
	2 R \, \sin \biggl( \frac{\Omega s}{2 \sqrt{1 + \Omega^2 R^2}} \biggr)
			\, \cos \biggl( \frac{\Omega t_0}{2 \sqrt{1 + \Omega^2 R^2}} \biggr), \\
& s =
	2 R \, \sin \biggl( \frac{\Omega t_0}{2 \sqrt{1 + \Omega^2 R^2}} \biggr)
			\, \cos \biggl( \frac{\Omega s}{2 \sqrt{1 + \Omega^2 R^2}} \biggr).
\end{aligned}
\right.
\end{equation}
These new equations must be solved together for $t$ and $s$ as
functions of $t_0$. The resulting distribution for differential aging
is shown in Fig.\ \ref{fig:twinrot}, and it is a smoother version of
the distribution that we obtained for the linear geometry of App.\
\ref{app:twins} (see Fig.\ \ref{fig:lapses}). Interestingly, if we set
\begin{equation}
\{ \tilde{t}, \tilde{s}, \tilde{t}_0 \} =
\frac{\Omega}{\sqrt{1 + \Omega^2 R^2}} \, \{ t, s, t_0 \},
\end{equation}
and multiply Eqs.\ \eqref{eq:finaltwin} by $\Omega / \sqrt{1 +
\Omega^2 R^2}$, we find that the solutions $\tilde{t}(\tilde{t}_0)$
and $\tilde{s}(\tilde{t}_0)$ depend on the product $\Omega R$, but not
on $\Omega$ and $R$ separately. This means that in these units, where
the total elapsed Lorentz time is just $2 \pi$, the shape of curve
that describes the aging distribution depends on Ulysses' absolute
velocity ($\Omega R = v / \sqrt{1 - v^2}$), but not on the radius and
angular frequency of his helix.
\end{appendix}

\providecommand{\bysame}{\leavevmode\hbox to3em{\hrulefill}\thinspace}

\end{document}